\documentstyle[prd,aps,preprint,epsfig]{revtex}

\tighten

\begin{document}
\draft
 
\pagestyle{empty}

\preprint{
\noindent
\hfill
\begin{minipage}[t]{3in}
\begin{flushright}
LBNL--47460 \\
February 2001
\end{flushright}
\end{minipage}
}

\title{AN OBVIOUS ISOSPIN BREAKING CORRECTION TO $\epsilon'$ OF
KINEMATICAL ORIGIN}

\author{
Mahiko Suzuki
}
\address{
Department of Physics and Lawrence Berkeley National Laboratory\\
University of California, Berkeley, California 94720
}


\maketitle

\begin{abstract}

Isospin breaking correction to the $\Delta I = \frac{1}{2}$ decay of
$K\rightarrow\pi\pi$ generates a large enough contribution to 
$\epsilon'$ through an induced $\Delta I = \frac{3}{2}$ amplitude 
in $K_L$ decay. Aside from the $\pi$-$\eta$ and $\pi$-$\eta'$ 
mixing contributions, there is a correction of kinematical origin due to 
the final-state $\pi^{\pm}$-$\pi^0$ mass difference, which is 
unambiguously calculable from the low-energy off-shell behavior of 
the $K\rightarrow\pi\pi$ amplitude. This correction to $\epsilon'$ reduces 
the isospin breaking parameter $\Omega_{IB}$ by 0.06, which is nearly one 
half of the $\pi$-$\eta$ mixing effect computed with chiral Lagrangians
to $O(p^2)$.

\end{abstract}
\pagestyle{plain}
\narrowtext

\setcounter{footnote}{0}

\section{Introduction}

The $\epsilon'$ parameter measures the CP phase difference between
the $\Delta I = \frac{1}{2}$ and $\Delta I = \frac{3}{2}$ amplitudes
in $K^0$ decay:
\begin{equation}
        \epsilon' = \frac{-i}{\sqrt{2}}e^{i(\delta_2-\delta_0)}\omega
          \biggl(\frac{{\rm Im}A_0}{{\rm Re}A_0}
          -\frac{{\rm Im}A_2}{{\rm Re}A_2}\biggr),
                   \label{epsilon'}
\end{equation}
where $A_{0,2}$ are the $K\rightarrow\pi\pi$ amplitudes into $I_{\pi\pi}=
0$ and $I_{\pi\pi}=2$. In the Standard Model with the conventional CP phase
assignment to quark mixing, the penguin decay is practically the entire 
source of ${\rm Im}A_0$. 
Because of the $\Delta I = \frac{1}{2}$ enhancement, the isospin breaking
correction to the penguin decay is a major $\Delta I = \frac{3}{2}$ 
contribution to $\epsilon'$. The $\pi$-$\eta$ and $\pi$-$\eta'$ mixing
contributions to the isospin breaking have been computed\cite{early}.

The isospin breaking effect is parametrized by
\begin{equation}
 \Omega_{IB} = \frac{{\rm Im}A_2^{IB}}{\omega{\rm Im}A_0}, \label{Omega}
\end{equation}
where  $\omega ={\rm Re} A_2/{\rm Re} A_0\simeq 1/22$, and
$A_2^{IB}$ is the $\Delta = \frac{3}{2}$ amplitude induced
by the electromagnetic or the $u$-$d$ quark mass difference correction to the
$\Delta I = \frac{1}{2}$ amplitude.  The isospin breaking parameter 
$\Omega_{IB}$ enters $\epsilon'$ as
\begin{equation}
        \epsilon' = \frac{-i}{\sqrt{2}}e^{i(\delta_2-\delta_0)}\omega
 \biggl[\frac{{\rm Im}A_0}{{\rm Re}A_0}(1-\Omega_{IB})
 -\frac{{\rm Im}A'_2}{{\rm Re}A_2}\biggr],
\end{equation}
where ${\rm Im}A'_2= {\rm Im}(A_2-A_2^{IB})$.
 The contribution of the $\pi$-$\eta$ mixing to $\Omega_{IB}$ is 
\begin{equation}
\Omega_{IB}^{\pi\eta} = 0.13 \label{pi-eta}
\end{equation} 
according to the calculation of $O(p^2)$ in chiral
Lagrangian\cite{early,Ecker}. The recent calculation to $O(p^4)$ raised the
value to $0.16\pm 0.03$\cite{Ecker}. After including $\pi$-$\eta'$ mixing,
$\Omega_{IB}^{\pi\eta+\pi\eta'}=0.25\pm 0.08$ has also been
quoted\cite{early}.

    The purpose of this short paper is to point out that there is one
obvious isospin breaking correction to the $\Delta I = \frac{1}{2}$ 
amplitude which arises from the $\pi^{\pm}$-$\pi^0$ mass difference 
of the final pions. This is an effect of $O(p^2)$ in the momentum 
expansion and numerically as large as one half of Eq. (\ref{pi-eta}) 
with the opposite sign. Considering its significance in testing the 
Standard Model with $\epsilon'$, we wish to call attention 
to this obvious correction.

\section{Isospin breaking by external pion mass difference}   

 The isospin structure of the $K\rightarrow\pi\pi$ decay amplitudes is
parametrized as
\begin{eqnarray}
  A(K^0\rightarrow\pi^+\pi^-) 
           &=& A_0(p_K^2, p_{\pi^{\pm}}^2, p_{\pi^{\pm}}^2) 
   + \frac{1}{\sqrt{2}}A_2(p_K^2, p_{\pi^{\pm}}^2, p_{\pi^{\pm}}^2) , 
                          \nonumber \\
  A(K^0\rightarrow\pi^0\pi^0) 
           &=& A_0(p_K^2, p_{\pi^0}^2, p_{\pi^0}^2) 
   - \sqrt{2} A_2(p_K^2, p_{\pi^0}^2, p_{\pi^0}^2). \label{isospin}
\end{eqnarray}
The amplitude $A_0$ in Eq. (\ref{isospin}) for $I=0$ actually hides 
an $I=2$ component through the external pion mass dependence.
The external four-momentum dependence of the $A_0$ amplitude has 
been well known to $O(p^2)$:
\begin{equation}
 A(p_K^2, p_a^2, p_b^2) = \frac{1}{2}A'(0)(2p_K^2-p_a^2-p_b^2)
  +O(p^4),     \label{momentumdep}
\end{equation}
where $A'(0)$ is a constant, and $p_K$ and $p_{a,b}$ denote momenta
of $K^0$ and two final pions. This robust external momentum dependence is
a consequence of SU(3) symmetry of strong interaction and charge
conjugation property of the parity-violating nonleptonic
decay interaction \cite{Gell-Mann}, though it is more often discussed
with chiral symmetry nowadays. The $\pi^{\pm}$-$\pi^0$ mass difference of 
the final pions in Eq. (\ref{momentumdep}) generates an effective 
$\Delta I = \frac{3}{2}$ amplitude and contributes to $\Omega_{IB}$
through ${\rm Im}A_2$. To our surprise, this correction has not been 
counted in literature.

On the mass shell the $A_0$ amplitude of Eq. (\ref{momentumdep}) has
the external mass dependence to $O(m_P^2)$, 
\begin{eqnarray}
 A_0(K^0\rightarrow\pi^+\pi^-) &=& A'(0)(m_{K^0}^2-m_{\pi^{\pm}}^2), 
                                 \nonumber\\
 A_0(K^0\rightarrow\pi^0\pi^0) &=& A'(0)(m_{K^0}^2-m_{\pi^0}^2).
                    \label{expansion}
\end{eqnarray}
A $\Delta I = \frac{3}{2}$ amplitude emerges from the difference
$m_{\pi^{\pm}}^2-m_{\pi^0}^2$ in Eq. (\ref{expansion}). Define
the induced $\Delta I = \frac{3}{2}$ amplitude as 
\begin{equation}
     A_2^{IB} \equiv -\frac{\sqrt{2}}{3}A'(0)(m_{\pi^{\pm}}^2-m_{\pi^0}^2).
              \label{IB}
\end{equation}
Moving $A_2^{IB}$ from $A_0$ to $A_2$, we can rewrite Eq. (\ref{isospin}) 
up to the $O(p^4)$ correction as
\begin{eqnarray}
 A(K^0\rightarrow\pi^+\pi^-) &=& A'(0)(m_{K^{0}}^2-\langle m_{\pi}^2\rangle) 
     + \frac{1}{\sqrt{2}}(A_2 + A_2^{IB}), \nonumber \\
  A(K^0\rightarrow\pi^0\pi^0) &=& A'(0)(m_{K^0}^2-\langle m_{\pi}^2\rangle) 
     - \sqrt{2} (A_2 + A_2^{IB}),
\end{eqnarray}
where $\langle m_{\pi}^2\rangle =\frac{1}{3}(2m_{\pi^{\pm}}^2+m_{\pi^0}^2)$.

Substituting the imaginary part of Eq. (\ref{IB}) in Eq. (\ref{Omega}), 
we obtain the contribution of ${\rm Im}A_2^{IB}$ to $\Omega_{IB}$ as
\begin{eqnarray}
  \Delta\Omega_{IB} &=& -\frac{\sqrt{2}}{3\omega}\biggl(
  \frac{m_{\pi^{\pm}}^2-m_{\pi^0}^2}{m_{K^0}-\langle m_{\pi}^2\rangle}
 \biggr),                \nonumber \\
                     &=& - 0.058. \label{numerical}
\end{eqnarray}
This is the external mass difference contribution to $\Omega_{IB}$ 
to $O(p^2)$ of $A_0$. The sign is opposite to the $\pi$-$\eta$ and 
$\pi$-$\eta'$ contributions, and the magnitude is nearly one half 
of the $O(p^2)$ contribution of $\pi$-$\eta$ mixing\cite{early,Ecker},
\begin{eqnarray}
     \Omega_{IB}^{\pi\eta}&=&\frac{m_d-m_u}{3\sqrt{2}(m_s-m_{u,d})}, 
                     \nonumber \\
                &=& 0.13,
\end{eqnarray} 
and more than a third of $\Omega_{IB}^{\pi\eta}= 0.16\pm 0.03$\cite{Ecker}
which includes $O(p^4)$.

\section{Discussion}

It is obvious that this pion mass difference contribution is not counted 
in the $\pi$-$\eta$ and $\pi$-$\eta'$ mixing calculation. It is purely
kinematical in origin. Our number is at the level of $O(p^2)$ in the 
language of chiral Lagrangian expansion. While the $\pi$-$\eta$ mixing 
correction is fairly clean to $O(p^2)$, the $O(p^4)$ correction contains
more dynamical uncertainties. In our calculation we have ignored an 
explicit SU(3) breaking in $A_0(p_K^2,p_a^2,p_b^2)$ of Eq. 
(\ref{momentumdep}).
This is the only possible source of uncertainty involved in
Eq. (\ref{numerical}). We make a remark on it.

The $s$-$u/d$ quark mass difference in internal lines generates 
an SU(3)-breaking $A_0$ term that does not vanish in the soft meson 
limit. In chiral Lagrangians, one quark mass insertion does not generate
a nonderivative term for the $\Delta I = \frac{1}{2}$ decay
since ${\rm tr}(\lambda_6M_qU^{\dagger})$ can be diagonalized away. 
Therefore there is no correction of $O(m_P^2)$ to our result.
The internal quark mass correction is of $O(m^2_P)\times O(p^2)$, which 
is the same as $O(p^4)$ on the mass shell of $K\rightarrow\pi\pi$. 
An explicit computation of $O(p^4)$ to one-loop was made with the kaon 
off mass shell, while keeping the pion masses on shell and 
degenerate\cite{Pallante}. In this calculation 
the explicit SU(3) breaking of the quark mass insertion manifests itself 
in the terms proportional to $m_K^2-m_{\pi}^2$ instead of $p_K^2-m_{\pi}^2$.
This SU(3) breaking turns out to be very small primarily because of 
the loop factor $1/(4\pi f_{\pi})^2$. Therefore the next-order correction
to Eq. (\ref{numerical}) is expected to be small. If one wishes to compare 
our result with the $O(p^4)$ result of $\Omega_{IB} (=0.16\pm 0.03)$ 
from chiral Lagrangians, one had better compute for consistency the $A_0$
amplitude with the pions off mass shell and include the next order
terms of $(m_K^2-m_{\pi}^2)\times O(p^2)$ in Eq. (\ref{numerical}).
Though it is small in magnitude, the next-order correction to 
Eq. (\ref{numerical}) contains dynamical uncertainty. 

\acknowledgements

This work was supported in part by the Director, Office of Science, 
Division of High Energy and Nuclear Physics, of the
U.S.  Department of Energy under Contract DE--AC03--76SF00098 and in
part by the National Science Foundation under Grant PHY--95--14797.

 
\end{document}